\documentclass[conference]{IEEEtran}

\usepackage[utf8]{inputenc} 	
\usepackage[T1]{fontenc}
\usepackage{microtype}
\usepackage{amsmath}
\usepackage{amssymb}
\usepackage{amsfonts}
\usepackage{amsthm}
\newtheorem{lem}{Lemma}

\usepackage[pdftex]{graphicx}

\usepackage{siunitx}
\DeclareMathOperator*{\erf}{erf}

\usepackage[caption=false,font=footnotesize]{subfig}

\usepackage[final=true]{hyperref}

\usepackage{tikz}

\hypersetup{
	pdfauthor = {Ankit Kaushik et al.},
	pdftitle = {Crowncom 2014},
	pdfsubject = {Crowncom 2014},
	pdfcreator = {PDFLaTeX with hyperref package},
        pdfproducer = {PDFLaTeX}
}
\title{On the Deployment of Cognitive Relay as Underlay Systems}
\author{Ankit Kaushik, M. Rehan Raza, Friedrich K. Jondral \\
        Communications Engineering Lab, Karlsruhe Institute of Technology (KIT) \\
        \{\href{mailto:Ankit.Kaushik@kit.edu}{Ankit.Kaushik}, 
        \href{mailto:Friedrich.Jondral@kit.edu}{Friedrich.Jondral}\}@kit.edu,
	\href{mailto:Muhammad.Raza@student.kit.edu}{Muhammad.Raza@student.kit.edu}}
\begin{document}
%
\maketitle
\thispagestyle{empty}
\pagestyle{empty}
\begin{abstract}
The objective of this paper is to extend the idea of Cognitive Relay (CR). CR, as a secondary user, follows an underlay paradigm to endorse secondary usage of the spectrum to the indoor devices. To seek a spatial opportunity, i.e., deciding its transmission over the primary user channels, CR models its deployment scenario and the movements of the primary receivers and indoor devices. Modeling is beneficial for theoretical analysis, however it is also important to ensure the performance of CR in a real scenario. We consider briefly, the challenges involved while deploying a hardware prototype of such a system.  
\end{abstract}
\section{Introduction}

It has been one and half decade since Mitola's idea of embedding cognition into the radio \cite{mitola}. One way to deploy cognition is through dynamic access to an under utilized spectrum. This resolves the problem of spectrum scarcity. The cellular operators are faced with a recurring challenge of reduced coverage and capacity within their network. Following the trend of growing mobile services, the situation will worsen in the upcoming years \cite{Ericsson11}. To some extent, the state of the art small cell stations, namely femto stations and relay stations, are deployed to solve the coverage and capacity issues. However, the stations within a macro or micro cell still share a limited spectrum, which increases complexity due to co-tier interference. Additionally, to combat cross-tier interference, these small stations have to coordinate with their immediate macro or micro station \cite{Andrews12}. \\
As a result, major applications or standards like IEEE 802.22 \cite{url:802.22} are emerging, which claim an efficient utilization of spectrum. However, due to their large coverage requirement, their access is rather static or limited to white spaces only. 
Moreover, development of such standards demands a long development phase, hence, field testing using a hardware prototype to validate a certain functionality is delayed. Considering this, we introduce a dynamic application, where a cellular operator \textit{offloads} its bandwidth demands over the spatially unused bands. These bands are not owned by the operator and are acquired cognitively. Considering heterogeneous deployment with variable propagation distance, smaller cells show a great potential for offloading. Thereby, cellular operators can offload users active inside small cells to spatially unused bands and ensure capacity enhancement \cite{Qualcomm13}. Here, capacity is defined as the number of active users per unit area. \\ 
Goldsmith \textit{et al.} \cite{Goldsmith09} describe different paradigms for shared access: overlay, underlay and interweave. These paradigms have been well investigated using analytical expressions and simulations in the literature \cite{Ghasemi06, Gastpar07}. However, their behavior through hardware realization is still not properly understood. Rapid prototyping of such hardware that is capable of implementing auxiliary functions such as sensing and sharing-constraints, is another challenge. Besides ameliorating capacities, these paradigms should conform strictly to regulatory requirements. Hence, it is important to examine and validate their performance through a real deployment. 

\subsection*{Cognitive Relay}
\begin{figure}[!t]
	\centering
	\includegraphics[width = \columnwidth]{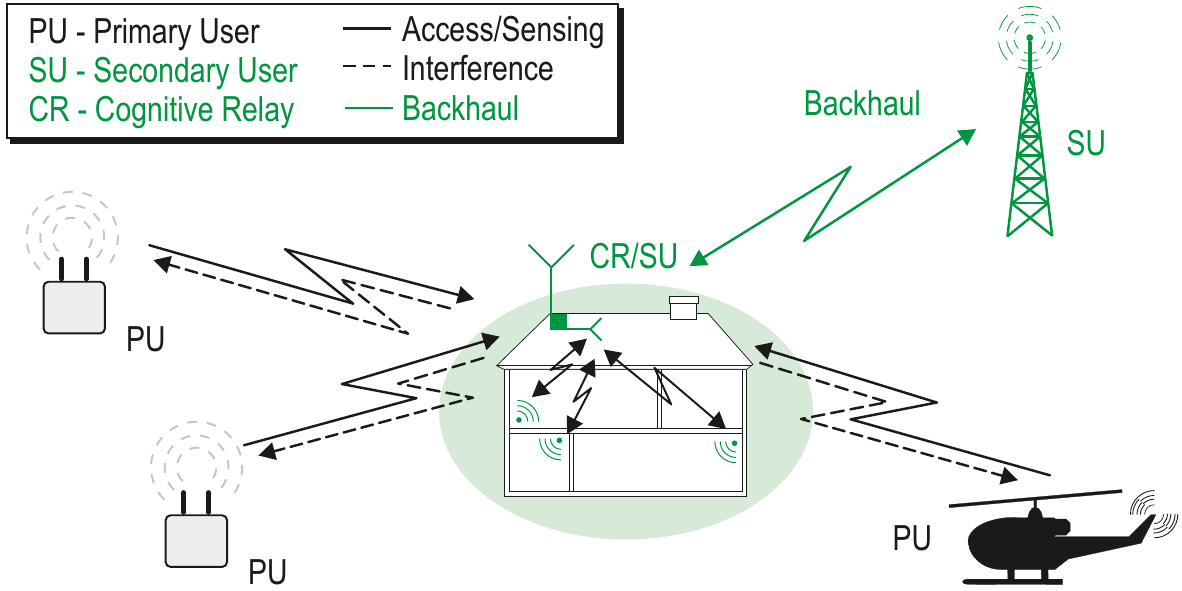}
	\caption{A scenario demonstrating the interaction between the PU and the CR. The CR senses PUs channels in the outdoor to provide a dynamic access to the devices operating indoor.}
	\label{fig:scenario}
\end{figure}

The network element responsible for realizing the concept of offloading is termed as Cognitive Relay (CR). CR introduced in \cite{Kaushik13} as a secondary user, provides access to indoor devices (IDs) over the unused spectrum. The knowledge of the unused spectrum is attained through sensing. An interweave scenario that finds temporal opportunities inside the primary user (PU) spectrum was demonstrated in \cite{Kaushik13}, however, in this paper the authors aim to deploy CR as underlay systems. \figurename~\ref{fig:scenario} depicts the operational scenario for the CR. 
CR links to the base station through a backhaul over an \textit{outdoor antenna}. Sensing and access over PUs' channels are done using an \textit{indoor antenna}. 
Following the guidelines from \cite{Goldsmith09}, we assume CR as an underlay system that executes transmissions only if it satisfies the sharing-constraints. \\ 
Through this paper, the authors like to extend the concept of CR as underlay system. We discuss the following challenges: modeling of propagation channel and sharing-constraints, from the hardware design perspective that are often neglected in theoretical studies. Far more important, as a validation process, we deploy the CR, ID and primary receiver (PR) in a real scenario and analyze the system performance through analytical expressions. Finally, we consider rapid prototyping and deployment of the system through a software defined radio architecture. 
\\
The rest of the paper is organized as follows: Section \ref{sec:sysm} explains the system model. It characterizes the channel models and the involved constraints. Section \ref{sec:meas} discusses the measurement results obtained after deploying the hardware. Section \ref{sec:conc} concludes the paper. The analytical expressions of the distribution function for interference at PR and capacity at ID are derived in appendix.
\section{System Model} \label{sec:sysm} 
A transmission over the PU channel, selected by CR for an indoor access between a CR and an ID, involves a time slotted half duplex communication. We consider that decision making occurs at CR. Thus, CR selects a PU channel for both uplink and downlink based on the sensing information available at CR and ID. This encourages IDs to sense the PU channels and feed their sensing information, that is energy measurements, back to the CR, e.g., over a Cognitive Pilot Channel \cite{Sallent09}. 
However to keep the discussion limited and yet tactful, we discuss only the downlink. Hence, sensing at IDs is not considered.\\ 
CR is obliged to avoid interference at the PRs and simultaneously ensure a Quality of Service (QoS) at the IDs. For the downlink, this entails the access channel $h_{\text{s}}$ between the CR and ID as \textit{indoor-indoor} link, and interference channel $h_{\text{p}}$ between CR and PR as \textit{indoor-outdoor} link, see \figurename~\ref{fig:scenario}. \\
Assuming channel reciprocity, the knowledge of channel states $h_{\text{p}}$ is acquired by sensing the beacons emitted by PR, and $h_{\text{s}}$ by observing the training sequence sent by ID. For fixed position of CR, ID and PR, the channel itself is static. It changes due to the mobility of PR and ID over the deployment scenario. Hence, CR should model the channel states to estimate the received power at PR and ID, and decide upon the PU channel for transmission which satisfies the sharing-constraints. This phenomenon is termed as \textit{spatial opportunity}. 
The performance of such a system is mainly governed by the accuracy of channel models \cite{Molisch09}. 
\subsection{Path loss model and large scale fading}
For the downlink access, the received signals at PR and ID encounter a distance dependent path loss. Additionally, due to the presence of different number of walls in between, PR and ID also witness shadowing. The mean value of path loss is determined using the log-distance model 
\begin{equation}
\label{eq:PL}
\text{PL}(d) = \text{PL}(d_0) + 10n\log_{10}\left(\frac{d}{d_0}\right),
\end{equation}
where $\text{PL}(d)$ denotes the path loss (dB) at a distance $d$, $d_0$ is a reference distance and $n$ represents the path loss exponent. 
The log-distance model is compared with two standard path loss models, i.e. ITU-R \cite{ITU-R} and WINNER II \cite{Winner2}. 
\begin{equation}
\label{eq:PL_ITU}
\text{PL}(d) = 20 \log_{10} f + 10n \log_{10} d + L_{\text{floors}}(k) - 28, 
\end{equation}
illustrates the ITU-R model, where $f$ is the center frequency (MHz) and $L_{\text{floors}}(k)$ represents the total attenuation for $k$ floors of the building. For propagation inside the office buildings at $f$ = 2.4 GHz, $n$~=~3 is used \cite[Table 2]{ITU-R}. 
\begin{equation}
\label{eq:PL_WINNER}
\text{PL}(d) = 20 \log_{10} \left(\frac{f}{5}\right) + 36.8 \log_{10} d + n_\text{w}L_\text{w} + 43.8,
\end{equation}
presents WINNER II model for a scenario with rooms and corridors \cite[(6.4), (6.5)]{Winner2}, where $f$ is the center frequency (GHz), $L_\text{w}$ denotes the attenuation through walls of the building ($L_\text{w}$ = \SI{5}{dB} 
for thin walls and $L_\text{w}$ = \SI{15}{dB} for thick walls) and $n_\text{w}$ is the average number of walls between CR and ID.   \\ 
In order to describe the variation of path loss about the mean value, shadowing is taken into account. Due to shadowing, the path loss $\text{PL}(d)$ in dB scale follows a normal distribution \cite{Goldsmith05}. (\ref{eq:PL}) shows one-one mapping between $\text{PL}(d)$ and the path loss exponent $n$. Correspondingly, $n$ also follows a normal distribution with Cumulative Distribution Function (CDF) 
\begin{equation}
\label{eq:normal}
F_{n}(n) = \frac{1}{2} \left[ 1 + \erf \left( \frac{( n - \mu_n)}{\sqrt{2} \cdot \sigma_n} \right) \right],
\end{equation} 
where $\mu_n$ and $\sigma_n$ denote the mean and standard deviation of $n$ while $\erf(\cdot)$ represents the error function.
\subsection{Small scale fading}
The mean value of path loss and shadowing model the channel for large movements of PR and ID. The model parameters $\text{PL}(d)$ and $n$, according to (\ref{eq:PL}), depend mainly on the deployment scenario. For movements $< 10\lambda$ \cite{Rappaport92}, $\text{PL}(d)$ and $n$ are considered to be correlated, hence, assumed to be constant. Due to small movements $\approx \frac{\lambda}{2}$ inside a circular region $\mathbb{R}$ of radius 10$\lambda$, $\text{PL}(d)$ varies due to multipath or small scale fading. \\ 
To utilize spatial opportunities efficiently, CR captures small movements of PR and ID, defined as snapshot. A snapshot corresponds to a spatial configuration of a PR and an ID. It includes their small movements inside $\mathbb{R}$, cf. \figurename~\ref{fig:deploymentScenario}. Another snapshot will result in a different spatial configuration of PR and ID around CR. For a fixed CR, PR and ID, the channel coefficients witness frequency-flat and slow fading. Due to the small movements of the PR and ID, the channel coefficients are modeled stochastically as small scale fading.\\ 
For a snapshot, CR seeks a spatial opportunity over a channel if it satisfies the following probabilistic constraints: interference constraint (IC) and capacity constraint (CC). IC is accomplished when the probability that the interference $I$ at PR is above a certain threshold $I_{\text{th}}$, is below $\epsilon_{\text{I,out}}$  
\begin{equation}
\label{eq:IC}
1 - F_{I}(I_{\text{th}}) = \mathbb{P} (I \ge I_{\text{th}}) \le \epsilon_{\text{I,out}},
\end{equation}
where $\epsilon_{\text{I,out}}$ is defined as interference outage. 
Similarly, CC is fulfilled when the probability that the capacity $C$ at ID is below a certain threshold $C_{\text{th}}$, is below $\epsilon_{\text{C,out}}$
\begin{equation}
\label{eq:CC}
F_C(C_{\text{th}}) = \mathbb{P} (C \le C_{\text{th}}) \le \epsilon_{\text{C,out}},
\end{equation}
where $\epsilon_{\text{C,out}}$ is the capacity outage. $I_{\text{th}}$, $\epsilon_{\text{I,out}}$, $C_{\text{th}}$ and $\epsilon_{\text{C,out}}$ are the design parameters with their values known at CR. In order to track the constraints analytically, it is required to characterize the distribution functions $F_I$ and $F_C$ in (\ref{eq:IC}) and (\ref{eq:CC}). 
\\
CR models the channel coefficients $h_\text{p}$, $h_\text{s}$, using Rayleigh or Nakagami-$m$ distribution for the indoor-indoor and indoor-outdoor links. In literature, Rayleigh distribution is mostly preferred due to its analytical tractability. In contrast to that, Nakagami-$m$ accounts for the severity in fading through $m$ parameter, thus, it is more applicable in indoor scenarios. For a fixed transmit power at CR, the received signal to noise power $\gamma$\footnote{Consider a transmission from a certain CR, then $\gamma$ is equivalent to signal to noise ratio (SNR) at ID and interference to noise ratio at PR. Therefore, the term SNR is not used to avoid confusion thereof. Moreover, the interference from unintended primary transmitters and other CRs at PR or ID is treated as white noise.} at PR or ID, corresponding to Rayleigh case, follows exponential distribution \cite{simon2005}
\begin{equation}
\label{eq:expo}
F_{\gamma}(\gamma) = 1 - e^{- \frac{\gamma}{\bar{\gamma}} }
\end{equation}
and for Nakagami-$m$, it follows Gamma distribution \cite{simon2005}
\begin{equation}
\label{eq:gamma}
F_{\gamma}(\gamma) = 1 - \frac{ \Gamma{ \left( m, m \frac{\gamma}{\bar{\gamma}} \right) }}{\Gamma(m)},
\end{equation} 
where $\bar{\gamma} = \mathbb{E}[\gamma]$ and $m$ denotes the shape parameter. $\Gamma(\cdot)$ and $\Gamma(\cdot,\cdot)$ are the complete and incomplete Gamma functions. The analytical expressions for $F_{I}$ and $F_C$ in (\ref{eq:IC}) and (\ref{eq:CC}), for the case when $h_{\text{p}}$ and $h_\text{s}$ are Rayleigh distributed, are presented in (\ref{eq:IC_distr_exp}) and (\ref{eq:CC_distr_exp}). When $h_{\text{p}}$ and $h_\text{s}$ are Nakagami-$m$ distributed, the $F_I$ and $F_C$ are given by (\ref{eq:IC_distr_gamma}) and (\ref{eq:CC_distr_gamma}). CR evaluates binary values for the individual constraints (\ref{eq:IC}) and (\ref{eq:CC}) corresponding to each snapshot. Finally, through an AND operation $(\cdot)$ over these constraints, CR decides to enable (= 1) or disable (= 0) the transmission over each channel. 
\section{Measurement Results and Discussions} \label{sec:meas}
\begin{figure}[!t]
        \centering
        \includegraphics[width = \columnwidth]{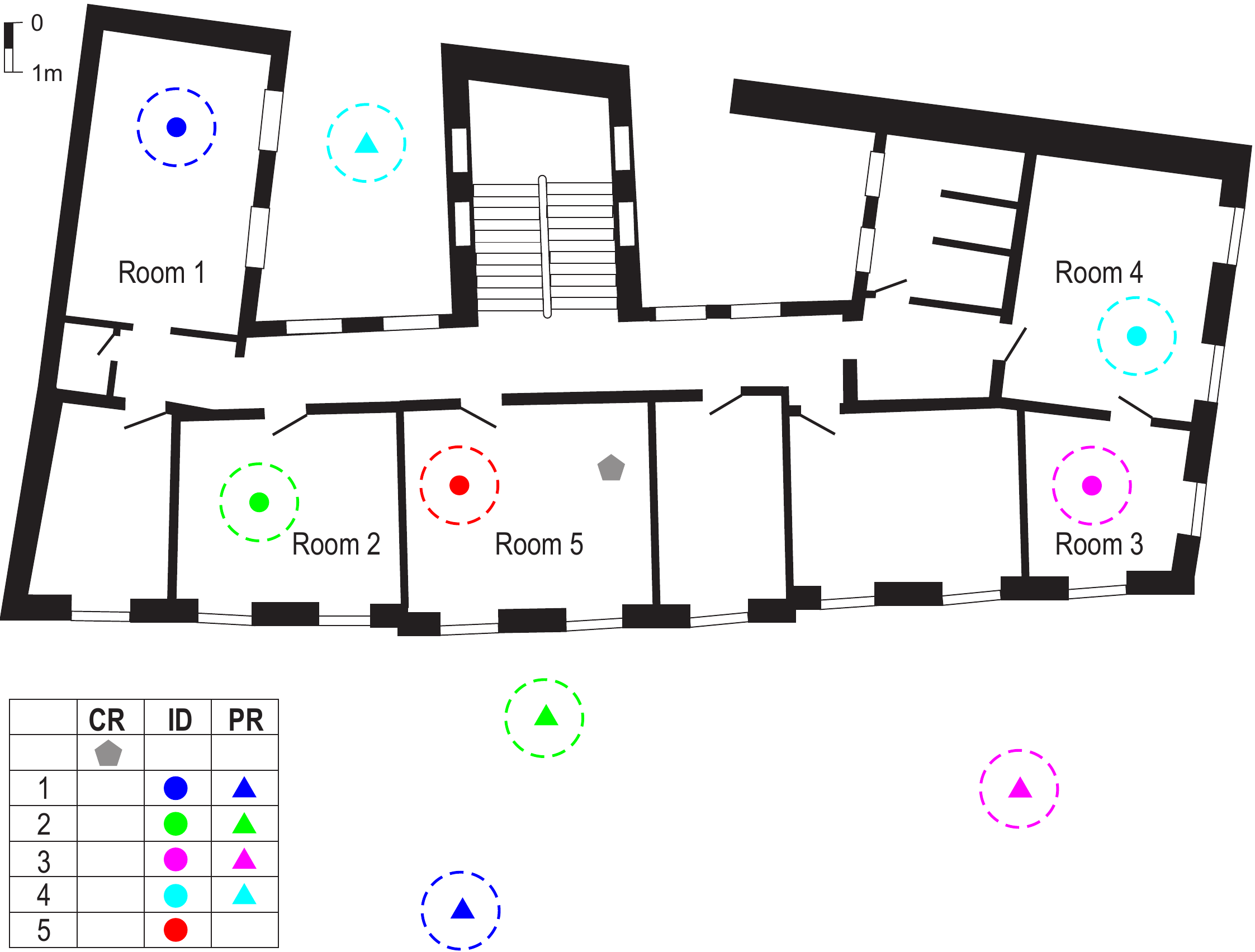}
        \caption{The deployment scenario for the CR including ID and PR at different spatial positions. The circle around the PR and ID positions represents $\mathbb{R}$, which illustrates their small movements.}
        \label{fig:deploymentScenario}
\end{figure}
\begin{figure*}[!t]
\centering
\subfloat[Least Squared Error fit for the log-distance model.]{\includegraphics[trim=1.5cm 0.4cm 1.45cm 1.2cm,clip=true,width=\columnwidth]{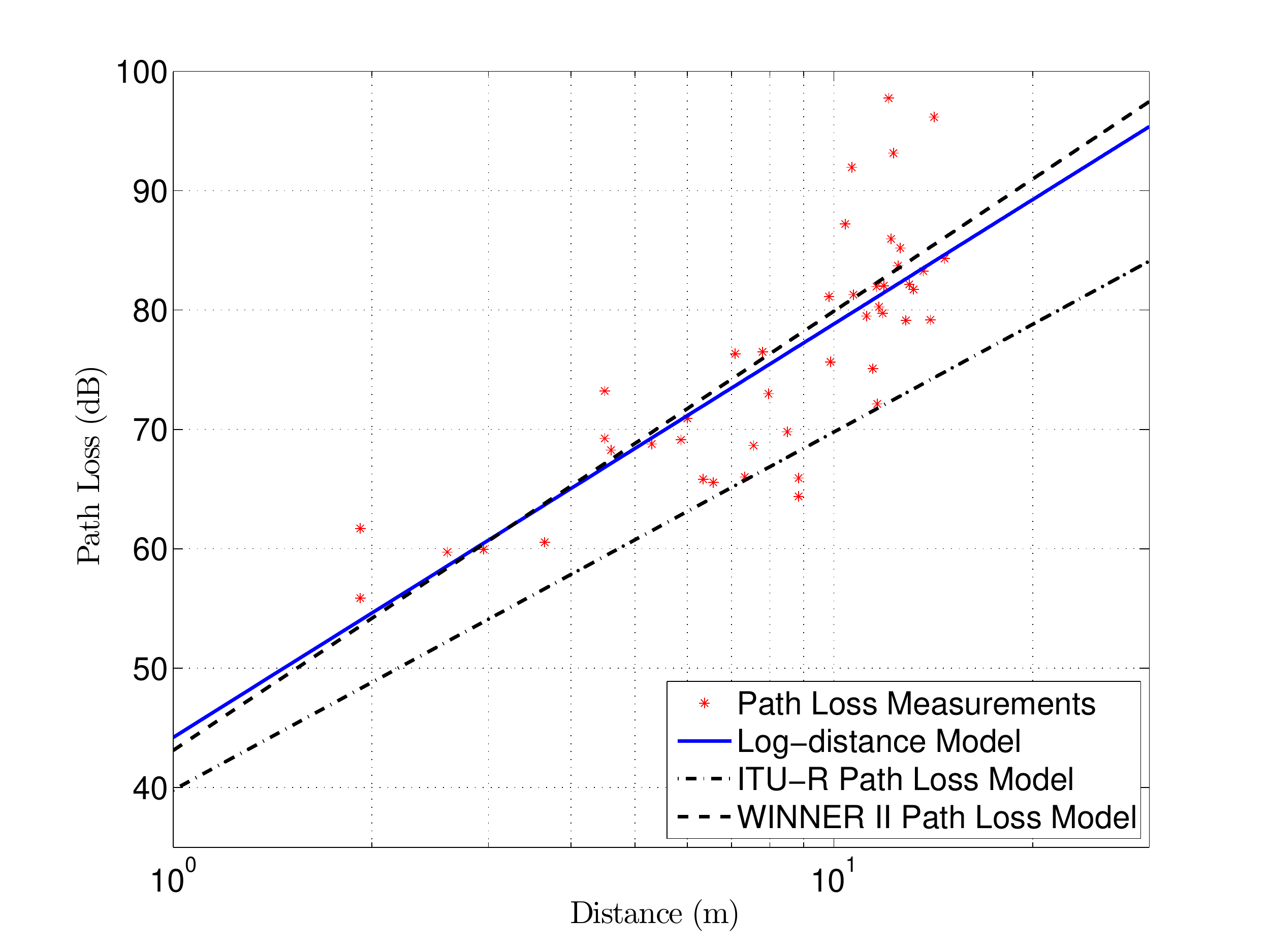}
\label{subfig:path_loss}}
\subfloat[Normal distribution $F_n$ describing shadowing.]{\includegraphics[trim=1.5cm 0.4cm 1.45cm 1.2cm,clip=true,width=\columnwidth]{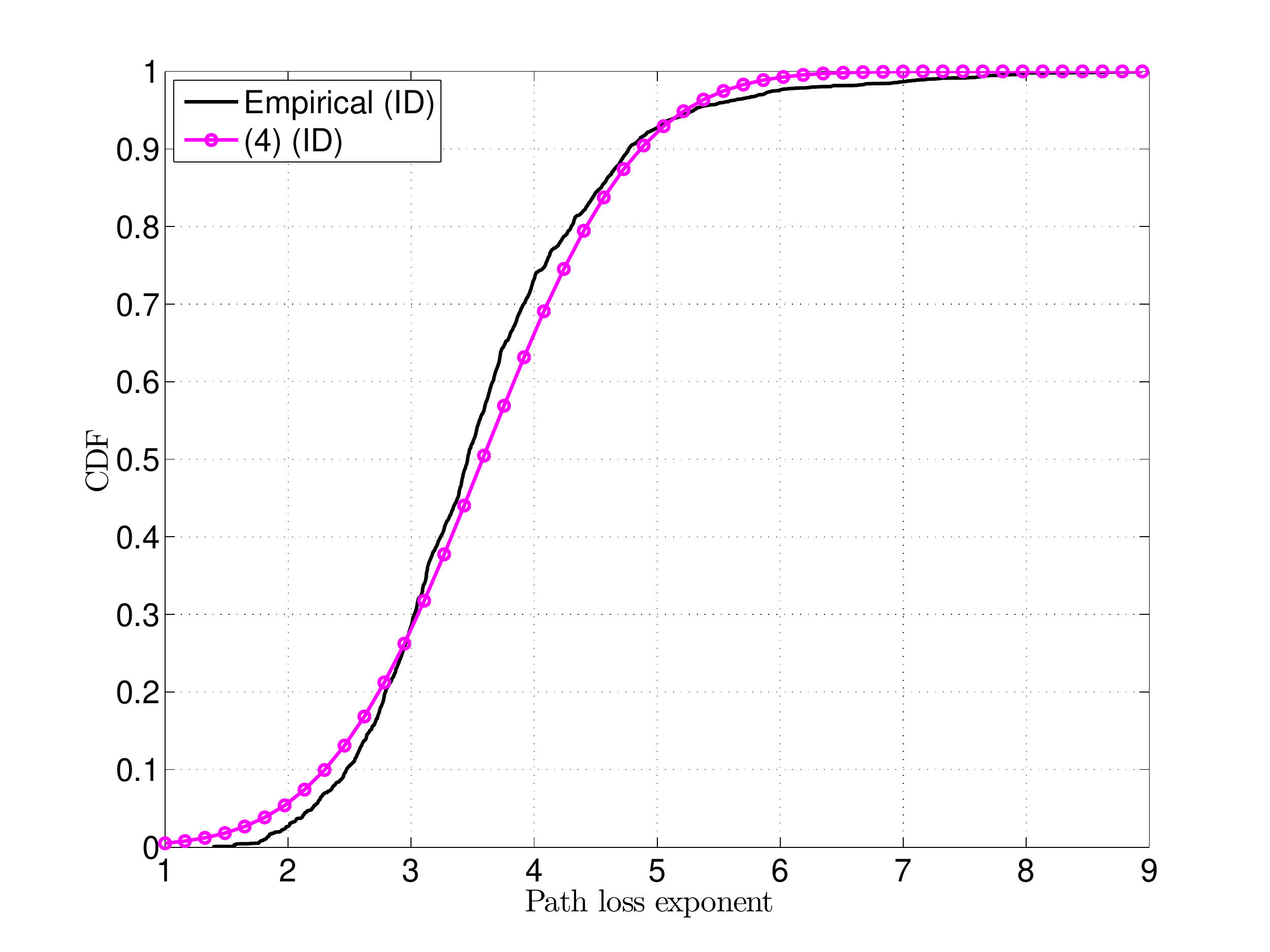}
\label{subfig:shadowing}}
\caption{Analytical expression for the log-distance model and the distribution function $F_n$ of the path loss exponent $n$ for indoor-indoor link compared with the empirical results. \figurename~\ref{subfig:path_loss} further depicts the comparison of the log-distance model to ITU-R and WINNER II path loss models.} 
\label{fig:pathloss_shadowing}
\end{figure*}

The last section dealt with the theoretical aspects of CR as underlay system. Here, we examine the operational perspective of its hardware prototype. \figurename~\ref{fig:deploymentScenario} presents the top view of the deployment scenario constituting CR, PR and ID. The number in the legend enumerates their configuration in the scenario. CR is mounted on the second floor of the building and IDs are also present at the same floor. The PRs are located outside the building at the ground level. 
\begin{figure}[!t]
	\centering
	\includegraphics[width = 0.8\columnwidth]{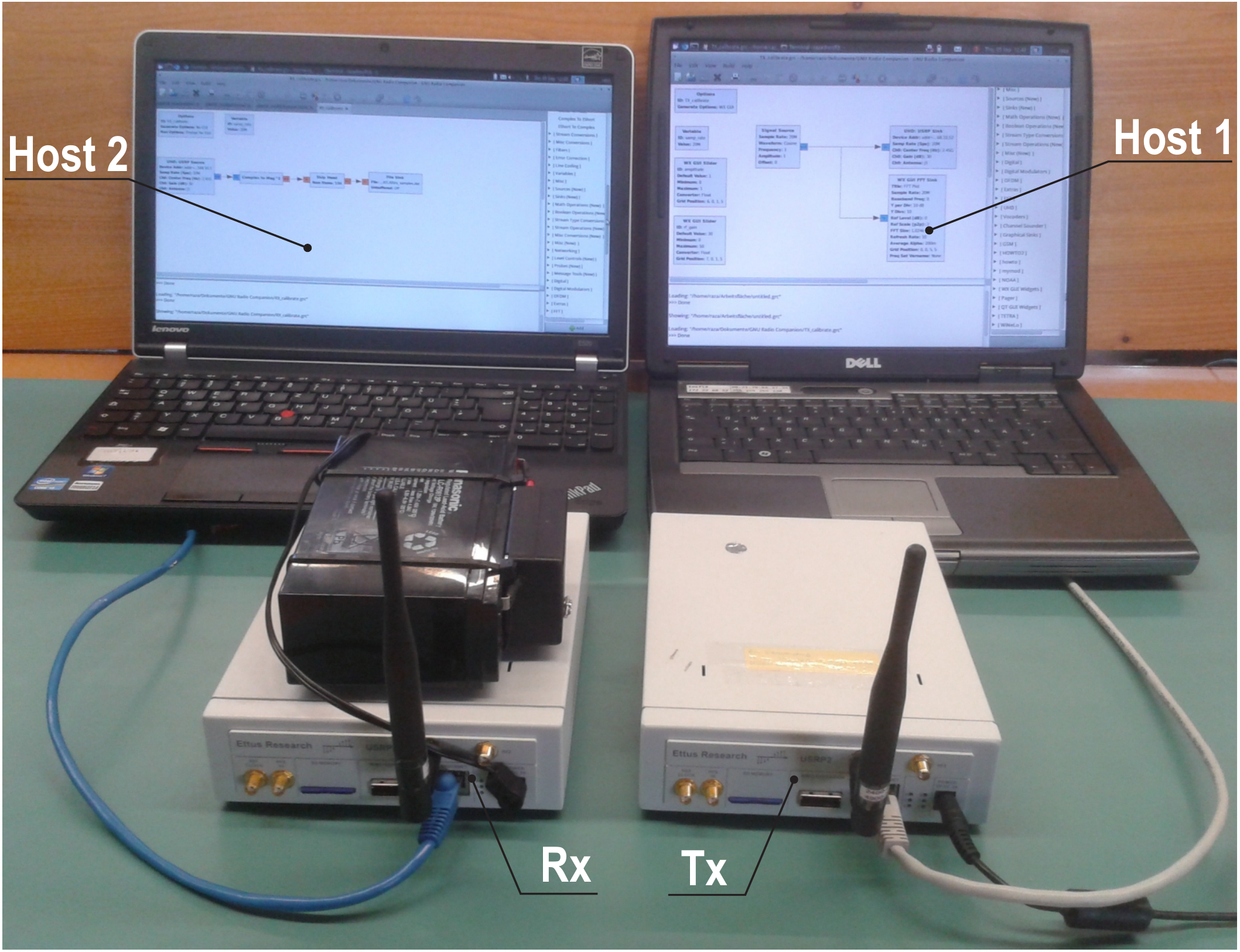}
	\caption{Hardware setup.}
	\label{fig:hw_setup}
\end{figure}
\\
To foster rapid prototyping of the auxiliary functions, 
a software defined platform is selected. The actors CR, PR and ID in \figurename~\ref{fig:deploymentScenario} are realized using Universal Software Radio Peripherals (USRPs) N200 \cite{url:usrp}. The host computers are connected to the USRPs via Ethernet cable, as shown in \figurename~\ref{fig:hw_setup}, to transfer digital data. The digital power corresponding to the transmitted and received samples in the USRPs is calibrated against the analog dBm power scale. This is done using R\&S FSL network analyzer on the transmitter side and R\&S 200A signal generator on the receiver side.
\\
In \figurename~\ref{fig:deploymentScenario}, indoor-indoor and indoor-outdoor links are separated by a boundary wall. Both the links are expected to follow a similar characterization for path loss and shadowing. Without loss of generality, for the path loss and shadowing, we consider only the indoor-indoor link. Unless explicitly specified, we consider a USRP as PR or ID, that transmits a sinusoidal signal at a constant transmit power = 10 dBm, and another USRP as CR, that evaluates the received power, thereby, estimating the channel states. Due to the availability of only one transmit-receive pair of USRPs, with no loss of generality, the measurements for PR to CR and ID to CR are discrete, and taken at different instants of time. In order to apply the independent property for the estimation of the model parameters, the measurements with a separation distance of $\approx 20 \lambda$ for shadowing, and $\approx \frac{\lambda}{2}$ for small scale fading are considered. \\ 
\figurename~\ref{subfig:path_loss} illustrates the path loss evaluated at CR for different ID positions. The model parameters in (\ref{eq:PL}), determined using Least Squared Error, are presented in TABLE \ref{tb:MSE}. $\widehat{\sigma}_{\text{PL}}$(dB) represents the standard deviation for the $\text{PL}(d)$. \figurename~\ref{subfig:path_loss} also compares the log-distance model with two standard models, i.e. ITU-R and WINNER II. The later model results in a closer approximation to the deployment scenario, as it provides an extra degree of freedom, which includes losses due to average number of walls $n_{\text{w}}$, consider (\ref{eq:PL_WINNER}). For our scenario, $\widehat{n}_{\text{w}} = 2.1$ is determined. \\
\figurename~\ref{subfig:shadowing} characterizes shadowing as normal distribution, as described in (\ref{eq:normal}). The model parameters $\widehat{\mu}_n$ and $\widehat{\sigma}_n$ for the path loss exponent are estimated using Maximum Likelihood Estimation (MLE), refer to TABLE \ref{tb:MSE}. Furthermore, the accuracy of this model is evaluated using Mean Squared Error (MSE).\\
Besides long term, CR response to the short term spatial opportunistic access is governed by PR and ID configuration. Based on the measurements for 4 PR and 5 ID positions, cf. \figurename~\ref{fig:deploymentScenario}, CR estimates the model parameters, corresponding to Rayleigh ($\widehat{\bar{\gamma}}$) and Nakagami-$m$ ($\widehat{\bar{\gamma}},\widehat{m}$) using MLE criterion. 
\figurename~\ref{subfig:indoor-outdoor} and \figurename~\ref{subfig:indoor-indoor} compare the analytical expressions of $F_I$ and $F_C$, presented in appendix, with the empirical CDFs. 
\begin{table}[!t]
\renewcommand{\arraystretch}{1.3}
\caption{Left: Parameters of the path loss modelled using log-distance model, Right: parameters and MSE of the shadowing modelled using normal distribution for indoor-indoor link}
\label{tb:MSE}
\centering
\begin{tabular}{c|c}
\hline
\bfseries Log-distance model (\ref{eq:PL}) & \bfseries Normal (\ref{eq:normal}) \\
{[$\widehat{\text{PL}}(d_0) \text{(dB)},\widehat{\sigma}_{\text{PL}} \text{(dB)}, \widehat{n}$]} & [MSE, $\mathcal{N} (\widehat{\mu}_n, \widehat{\sigma}_n)$] \\
\hline\hline
$[44.19,5.94,3.46]$  & $[1.70 \cdot 10^{-3},  \mathcal{N} (3.58, 1.00)]$ \\ \hline
\end{tabular}
\end{table}
\begin{table}
\renewcommand{\arraystretch}{1.3}
\caption{Parameters and MSE of the $F_{I}$ modeled using Rayleigh and Nakagami-$m$ distribution for different PR positions}
\label{tb:MSE_II}
\centering
\begin{tabular}{c||c|c}
\hline
\bfseries Outdoor & \bfseries Rayleigh (\ref{eq:expo}) & \bfseries Nakagami-$m$ (\ref{eq:gamma}) \\
\bfseries Position & [MSE, $\widehat{\bar{\gamma}}]$ & [MSE, $(\widehat{m}, \widehat{\bar{\gamma}}$)] \\
\hline\hline
PR1 & $[3.84 \cdot 10^{-4}, 2.66 \cdot 10^{2} ]$  & $[1.74 \cdot 10^{-4} , (1.13, 2.66 \cdot 10^{2})] $ \\ \hline
PR2 & $[2.75 \cdot 10^{-4}, 4.89 \cdot 10^{2}]$  & $[2.40 \cdot 10^{-4} , (0.98, 4.89 \cdot 10^{2})] $ \\ \hline
PR3 & $[2.75 \cdot 10^{-4}, 57.34]$  & $[2.31 \cdot 10^{-4} , (1.11, 57.34)] $ \\ \hline
PR4 & $[9.36 \cdot 10^{-4}, 94.20]$  & $[2.18 \cdot 10^{-4} , (1.25, 94.20)] $ \\ \hline
\end{tabular}
\end{table}
\begin{table}
\renewcommand{\arraystretch}{1.3}
\caption{Parameters and MSE of the $F_{C}$ modeled using Rayleigh and Nakagami-$m$ distribution for different ID positions}
\label{tb:MSE_IO}
\centering
\begin{tabular}{c||c|c}
\hline
\bfseries Indoor & \bfseries Rayleigh (\ref{eq:expo}) & \bfseries Nakagami-$m$ (\ref{eq:gamma}) \\
\bfseries Position & [MSE, $\widehat{\bar{\gamma}}]$ & [MSE, $(\widehat{m}, \widehat{\bar{\gamma}}$)] \\
\hline\hline
ID1 & $[1.11 \cdot 10^{-3}, 9.52 \cdot 10^{2}]$  & $[1.74 \cdot 10^{-4} , (1.23, 9.52 \cdot 10^{2})] $ \\ \hline
ID2 & $[1.60 \cdot 10^{-3}, 3.65 \cdot 10^{4}]$  & $[3.39 \cdot 10^{-4} , (1.28, 3.65 \cdot 10^{4})] $ \\ \hline
ID3 & $[5.79 \cdot 10^{-4}, 1.79 \cdot 10^{2}]$  & $[1.35 \cdot 10^{-4} , (1.17, 1.79 \cdot 10^{2})] $ \\ \hline
ID4 & $[9.35 \cdot 10^{-4}, 4.13 \cdot 10^{2}]$  & $[3.05 \cdot 10^{-4} , (1.16, 4.13 \cdot 10^{2})] $ \\ \hline
ID5 & $[7.54 \cdot 10^{-4}, 6.99 \cdot 10^{4}]$  & $[2.96 \cdot 10^{-4} , (1.23, 6.99 \cdot 10^{4})] $ \\ \hline
\end{tabular}
\end{table}
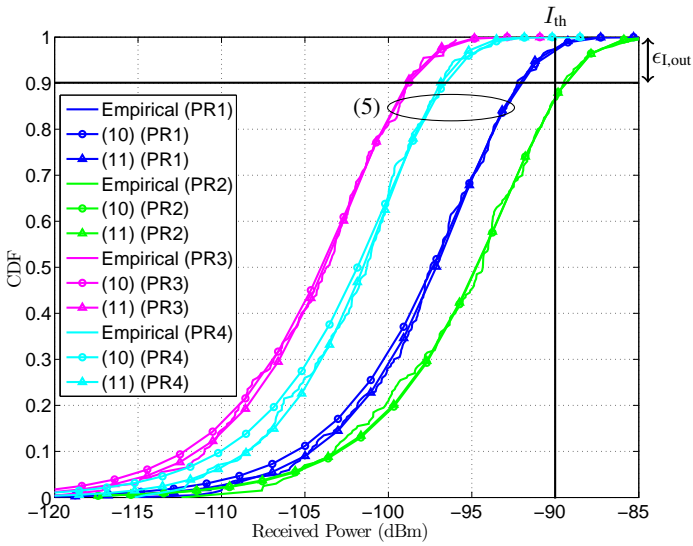
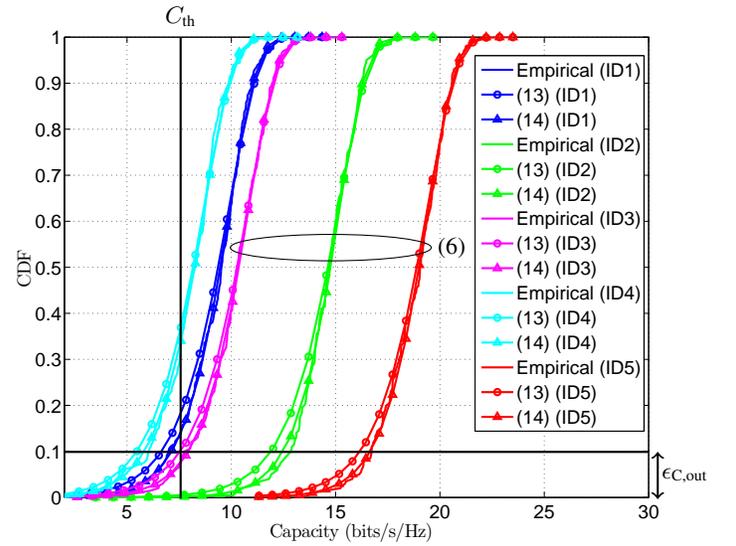
\begin{figure*}%
\centering
\subfloat[Interference distribution function at different PR positions.]{
\begin{tikzpicture}
\node[anchor=south west,inner sep=0] (image) at (0,0) 
{\includegraphics[trim=1.5cm 0.4cm 1.45cm 1.2cm,clip=true,width=\columnwidth]{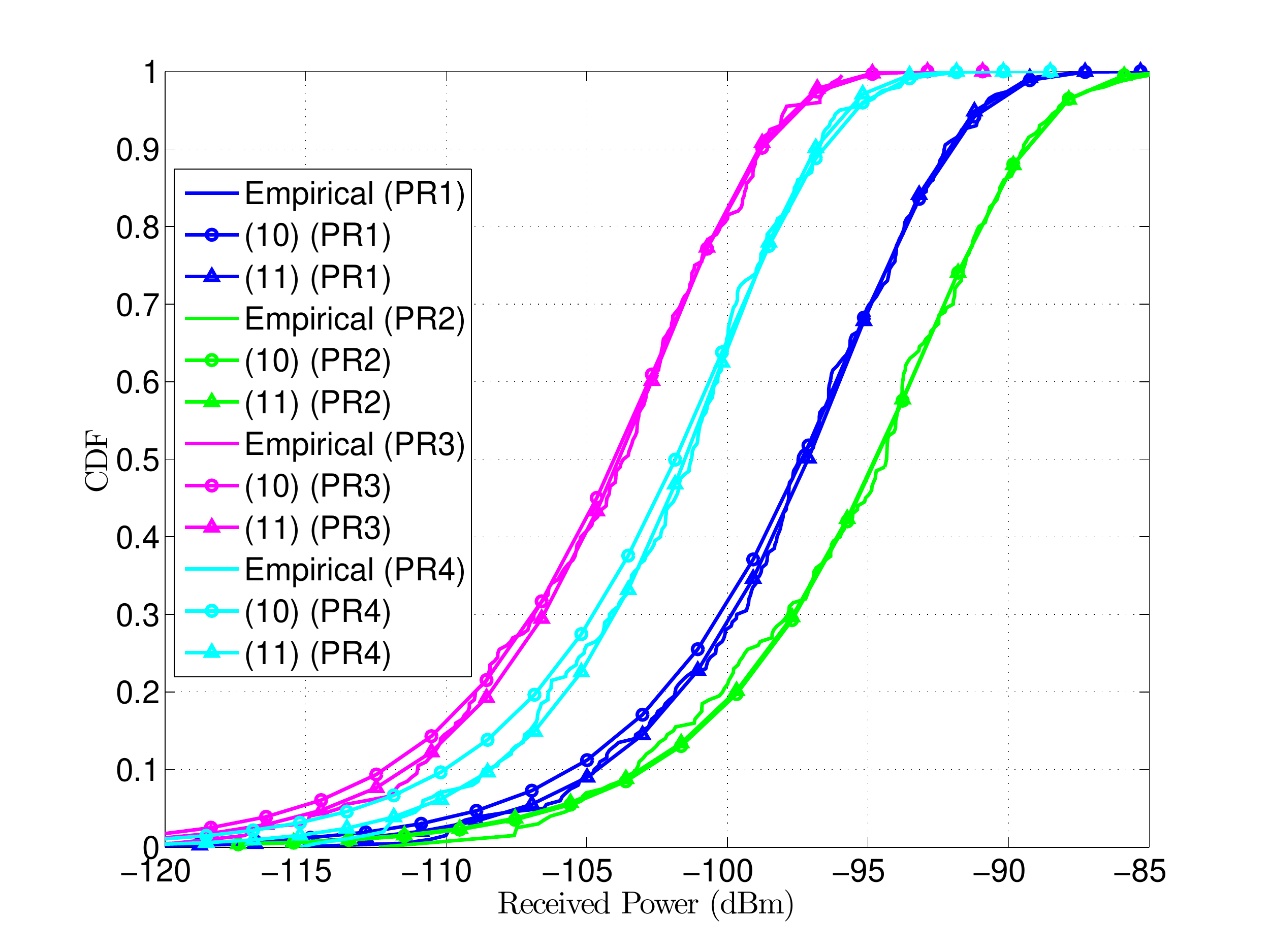}};
\begin{scope}[x={(image.south east)},y={(image.north west)}]
\draw  (0.832,0.99) node[above=-0.1pt, font=\small] {$I_{\text{th}}$};
\draw[black,thick,<->] (0.97,0.898) --  node[right=-2.2pt, font=\small] {$\epsilon_{\text{I,out}}$} (0.97,0.985);
\draw [thick] (0.831,0.985) -- (0.831,0.096);
\draw [thick] (0.08,0.898) -- (0.956,0.898);

\draw (0.675,0.85) ellipse(24pt and 5pt)  node[left=23pt, font=\small] {(5)};
\end{scope}
\end{tikzpicture}
\label{subfig:indoor-outdoor}}
\subfloat[Capacity distribution function at different ID positions.]{
\begin{tikzpicture}
\node[anchor=south west,inner sep=0] (image) at (0,0)
{\includegraphics[trim=1.5cm 0.4cm 1.45cm 1.2cm,clip=true,width=\columnwidth]{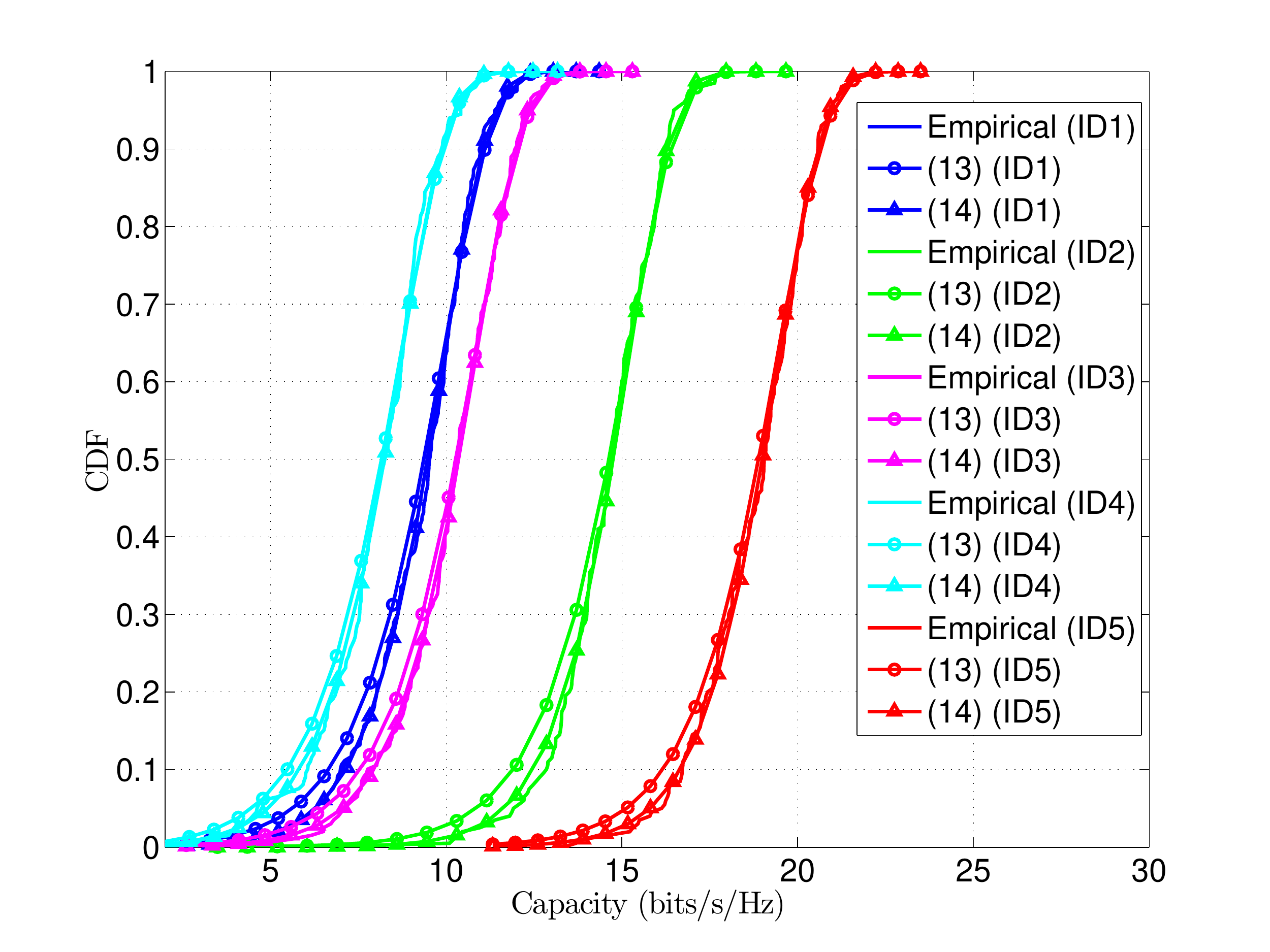}};
\begin{scope}[x={(image.south east)},y={(image.north west)}]
\draw  (0.255,0.99) node[above=-0.1pt, font=\small] {$C_{\text{th}}$};
\draw[black,thick,<->] (0.97,0.095) --  node[right=-2.2pt, font=\small] {$\epsilon_{\text{C,out}}$} (0.97,0.186);
\draw [thick] (0.255,0.985) -- (0.255,0.096);
\draw [thick] (0.08,0.186) -- (0.956,0.186);
\draw (0.48,0.58) ellipse (38pt and 5pt)  node[right=37pt, font=\small] {(6)};
\end{scope}
\end{tikzpicture}
\label{subfig:indoor-indoor}}
\caption{Analytical expressions of $F_I$ and $F_C$ rendered using Rayleigh and Nakagami-$m$ models compared with the empirical results. The number besides PR and ID depicts their position as illustrated in \figurename~\ref{fig:deploymentScenario}. The ellipse represents the ID and PR positions qualifying the corresponding constraints. Indoor position 3 (ID3) in \figurename~\ref{subfig:indoor-indoor} reveals the inaccuracy of the Rayleigh model. According to it, CR employing Rayleigh model prohibits its transmission to ID3 as it fails to satisfy the capacity constraint, which is inconsistent to the empirical values. However, this is not the case with Nakagami-$m$ model.}
\label{fig:smallscale}
\end{figure*}
\begin{table}[!t]
\renewcommand{\arraystretch}{1.3}
\caption{Cognitive relay implementing AND rule for different PR and ID positions}
\label{tb:BDecisions}
\centering
\begin{tabular}{c||c|c|c|c}
\hline
\bfseries & PR1 & PR2 & PR3 & PR4 \\
\hline\hline
ID1 & 1 $\cdot$ 0 = 0 & 0 $\cdot$ 0 = 0 & 1 $\cdot$ 0 = 0 & 1 $\cdot$ 0 = 0 \\ \hline
ID2 & 1 $\cdot$ 1 = 1 & 0 $\cdot$ 1 = 0 & 1 $\cdot$ 1 = 1 & 1 $\cdot$ 1 = 1 \\ \hline
ID3 & 1 $\cdot$ 1 = 1 & 0 $\cdot$ 1 = 0 & 1 $\cdot$ 1 = 1 & 1 $\cdot$ 1 = 1 \\ \hline
ID4 & 1 $\cdot$ 0 = 0 & 0 $\cdot$ 0 = 0 & 1 $\cdot$ 0 = 0 & 1 $\cdot$ 0 = 0 \\ \hline
ID5 & 1 $\cdot$ 1 = 1 & 0 $\cdot$ 1 = 0 & 1 $\cdot$ 1 = 1 & 1 $\cdot$ 1 = 1 \\ \hline
\end{tabular}
\end{table}
Once the model parameters for a snapshot are determined, CR enforces the interference and capacity constraints described in (\ref{eq:IC}) and (\ref{eq:CC}), using Nakagami-$m$ model since it yields smaller MSE. To define these constraints, the parameters values $I_{\text{th}} = -90$ dBm,  $\epsilon_{\text{I,out}} = 0.1$, $C_{\text{th}} = 7.5$ bits/sec/Hz and $\epsilon_{\text{C,out}} = 0.1$ were set, as shown in \figurename~\ref{subfig:indoor-outdoor} and \figurename~\ref{subfig:indoor-indoor}. However, these values can be adjusted depending on the system requirements. Table \ref{tb:BDecisions} presents the binary decisions attained after exercising the individual constraints. Table \ref{tb:BDecisions} also illustrates the final decisions, after the AND operation, for all 20 snapshots.  

\section{Conclusion} \label{sec:conc}
The paper introduces CR as underlay system, a cognitive radio application that realizes an indoor access through secondary usage. It discusses the challenges involved in constructing a hardware prototype of such a system. To specify the spatial opportunity for the downlink, we state the importance of the channels namely indoor-indoor and indoor-outdoor. Using software defined platform, we employ channel models to illustrate the PR and ID mobility in a real scenario. Thus, the performance of the system is determined by comparing empirical results with analytical expressions. Finally, the constraints are implemented for different snapshots to qualify the PU channels as spatial opportunities. The deployment aspects for the uplink, and modeling the mobility that entails large and small movements as Suzuki process are left for the future work. \\
The approach implemented in this paper, that considers constant transmit power at the CR, is static. However as a future step, the transmit power at the CR could be varied dynamically, in order to satisfy sharing constraints and utilize the spatial opportunities within PU spectrum more efficiently. 
\section*{Acknowledgment}
The authors thank the German Federal Ministry of Education and Research (BMBF) for partially funding their ongoing research under the grant 16BU1205 within the project Cognitive Mobility Radio (CoMoRa).
\section*{Appendix} \label{sec:appen}
\begin{lem}
\normalfont 
Distribution function of $F_I$ 
\label{lm:ID}
\begin{align}
\label{eq:IC_distr_derive}
F_I(I_{\text{th}}) &= \mathbb{P} (I \le I_{\text{th}}) = \mathbb{P} \left( \gamma \le \frac{I_{\text{th}}}{\sigma^2} \right) =  F_{\gamma} \left( \frac{I_{\text{th}}}{\sigma^2} \right),
\end{align}
where $\sigma^2$ represents the noise power at the PR. When $\gamma$ is exponentially distributed (\ref{eq:expo}), then $F_I$ in (\ref{eq:IC_distr_derive}) is obtained as
\begin{equation}
\label{eq:IC_distr_exp}
F_I(I_{\text{th}}) = 1 - e^{- \frac{({I_{\text{th}}}/{\sigma^2})}{\bar{\gamma}}}, 
\end{equation}
and for the case when $\gamma$ has Gamma distribution (\ref{eq:gamma}), $F_I$ in (\ref{eq:IC_distr_derive}) is determined as
\begin{equation}
\label{eq:IC_distr_gamma}
F_I(I_{\text{th}}) = 1 - \frac{ \Gamma{ \left( m, m \frac{({I_{\text{th}}}/{\sigma^2})}{\bar{\gamma}} \right) }}{\Gamma(m)}.
\end{equation}
\end{lem}
\begin{lem}
\normalfont 
Distribution function of $F_C$ 
\label{lm:CD}
\begin{align}
\label{eq:CC_distr_derive}
F_C(C_{\text{th}}) &= \mathbb{P} (C \le C_{\text{th}}) = \mathbb{P} \left( \log_{2} \left( 1 + \gamma \right) \le  C_{\text{th}} \right) \nonumber \\
\quad			   &= \mathbb{P} (\gamma \le 2^{C_{\text{th}}} - 1) = F_{\gamma}(2^{C_{\text{th}}} - 1).  
\end{align}
When $\gamma$ is exponentially distributed (\ref{eq:expo}), then $F_C$ in (\ref{eq:CC_distr_derive}) is obtained as
\begin{equation}
\label{eq:CC_distr_exp}
F_C(C_{\text{th}}) = 1 - e^{- \frac{(2^{C_{\text{th}}} - 1)}{\bar{\gamma}}}, 
\end{equation}
and for the case when $\gamma$ has Gamma distribution (\ref{eq:gamma}), $F_C$ in (\ref{eq:CC_distr_derive}) is determined as
\begin{equation}
\label{eq:CC_distr_gamma}
F_C(C_{\text{th}}) = 1 - \frac{ \Gamma{ \left( m, m \frac{(2^{C_{\text{th}}} - 1)}{\bar{\gamma}} \right) }}{\Gamma(m)}. 
\end{equation}
\end{lem}
\bibliographystyle{IEEEtran}
\bibliography{IEEEabrv,refs}
\end{document}